# Trap Dynamics and Conductivity Changes in AlGaN/GaN Heterostructures Under Ultrasonic Loading

Vladyslav Kaliuzhnyi[1, †], Mykola Tymochko[1], Oleksandr Gudymenko[1], Oleg Olikh[2], and Alexander Belyaev[1]

[1]V. E. Lashkaryov Institute of Semiconductor Physics, National Academy of sciences of Ukraine, Kyiv, 03028, Ukraine

[2]Taras Shevchenko National University, Kyiv, 01601, Ukraine

**Abstract:** AlGaN/GaN heterostructures are critical for high-power electronic devices but suffer from electron trapping at defects, limiting reliability. We investigate how ultrasonic vibrations affect electron transport in MOCVD-grown AlGaN/GaN heterostructures by combining temperature-dependent Hall effect and high-resolution X-ray diffraction measurements. Here we demonstrate that ultrasonic loading induces persistent acoustoconductivity and lattice parameter changes, attributed to acoustically driven rearrangement of metastable DX centers. This leads to increased carrier concentration and decreased mobility, reflecting defect state modulation by acoustic strain. These findings provide new insights into trap dynamics under dynamic deformation and suggest ultrasonic treatment as a potential approach to mitigate trapping effects, thereby enhancing the performance and reliability of GaN-based devices.

## 1. Introduction

Properties of the two-dimensional electron gas (2DEG) at the AlGaN/GaN interface have attracted significant attention due to the great potential of GaN-based devices for electronic applications, such as high-electron-mobility transistors (HEMTs),[1] as well as optoelectronic devices, including lasers and light-emitting diodes (LEDs).[2] AlGaN/GaN-based HEMTs are particularly promising for high-voltage, high-frequency, and high-power applications.[3,4] The outstanding intrinsic material parameters of these heterostructures enable their operation under extreme conditions. In particular, interfacial charges induced by piezoelectric and pyroelectric effects in this material system allow for control of the 2DEG electron concentration by adjusting the thickness and Al content of the AlGaN layer grown on the (0001) GaN surface, eliminating the need for modulation doping.[5] Despite these advantages, GaN HEMT technology still faces significant reliability challenges. A major issue affecting device performance is the phenomenon of electron trapping,[6] which leads to effects such as current dispersion or current collapse—reductions in drain current during high-voltage operation due to trapped charges.[7] These traps can be located on the surface of the HEMT structure or distributed within the AlGaN barrier or GaN buffer regions.[8] Among the various types of traps, deep donor defects known as DX centers are of particular interest due to their metastable nature and significant impact on carrier dynamics. However, the mechanisms by which these traps influence device performance, especially under dynamic conditions, are still not well understood. To address this knowledge gap, we investigate the influence of ultrasonic vibrations on the mobility and carrier concentration in AlGaN/GaN heterostructures, with particular attention to the role of electrically active traps such as DX centers in electron transport. Our results demonstrate that ultrasonic treatment can controllably modify transport properties by inducing trap restructuring, providing new insights into the mechanisms of trap-related phenomena and suggesting potential strategies to improve the reliability of GaN-based devices.

† Correspondence to: V. Kaliuzhnyi, Email: vladkaliuzh@gmail.com

## 2. Experiment

The structure in this study was grown by metalorganic chemical vapor deposition (MOCVD). A 5 nm AlN layer, commonly referred to as the "back barrier," was unintentionally doped and grown on (0001) GaN/sapphire template. This template served as the foundation for the HEMT heterostructure, which consisted of a 300 nm GaN buffer layer, a 20 nm $Al_{0.2}Ga_{0.8}N$ barrier layer, and a 5 nm GaN cap layer.

The structure was investigated with high-resolution X-ray diffraction (HRXRD) using PANalytical X'Pert Pro MRD XL diffractometer equipped with the $CuK\alpha_1$ radiation ($\lambda$ = 0.154056 nm), four-bounce (220) Ge monochromator, and three-fold (220) Ge analyzer. In order to characterize the layers, Hall effect under the van der Pauw configuration was performed using annealed at 450°C Indium contacts placed at the four corners of a square (6 × 6 $mm^2$). The experiments were conducted in a magnetic field of 0.47 T and over a temperature range of 80 to 315 K.

In this study, we employed a dynamic deformation method in which ultrasonic vibrations of a tunable frequency and adjustable amplitude were applied to the wafer. Ultrasonic vibrations are generated in the wafer using an external $LiNbO_3$ piezoelectric transducer. The vibrations propagate into the wafer from the transducer and form quasi-standing acoustic waves at resonance frequency. The specific power introduced into sample equals $P_{us} \sim 10^4$ W/m$^2$ at voltage of 15 V applied to transducer that corresponds deformation in acoustic wave $\varepsilon_{us} = (2P_{us}/\rho\upsilon^3)^{1/2} \sim 10^{-6}$ and strain of $\tau_{us} = (2\rho\upsilon P_{us})^{1/2} \sim 10^6$ N/m$^2$, where $\rho$ and $\upsilon$ are density and sound velocity, respectively. Longitudinal acoustic wave propagates along piezoactive [0001] direction. Thermoelectric cooler was used to stabilize temperature under applied voltage. Schematic of packaged AlGaN/GaN HEMT mounted on the acoustic unit is shown in Fig.1(a). Schematic of sample is shown at Fig.1(b).

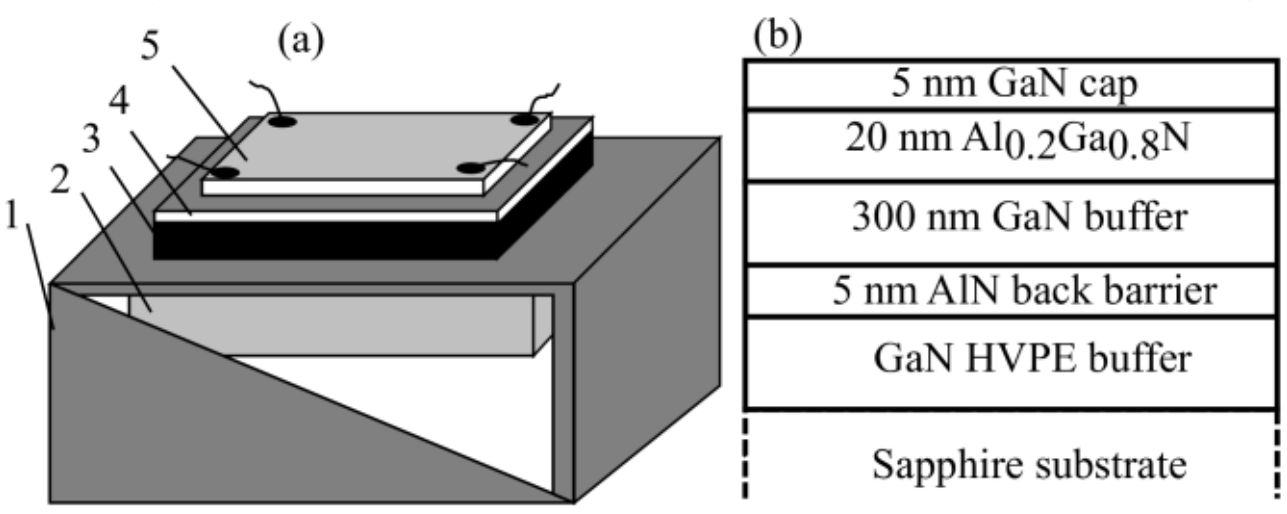

Fig. 1. Experimental setup (a) and sample layout (b). 1 – shield case, 2 – piezoelectric converter, 3 – acoustic medium (metal plate), 4 – mica foil substrate, 5 – sample.

## 3. Results and discussion

Temperature dependencies of carrier density and mobility for our sample, both with and without exposure to ultrasound, are shown in Fig. 2. To eliminate the influence of ambient illumination, all data were collected in the dark. Initially, we focus on the data obtained without ultrasound application. At high temperatures, the mobility decreases with increasing temperature, while at low temperatures it remains nearly constant, which is characteristic of a 2DEG system. However, the measured sheet carrier concentration at low temperatures is significantly higher than the value calculated for the specific AlGaN/GaN interface. This discrepancy may arise from the presence of free carriers in the GaN buffer layer, traps within the barrier, or differences in the surface states of the AlN barrier layer and the thickness of the GaN layer.[9] Additionally, thermally activated carrier generation—driven by polarization-induced charge modulation at the AlN/GaN interface, thermal ionization of shallow donors,[10] or the influence of the so-called "back barrier" (the AlN interlayer),[11]—could contribute to this behavior. However, no measurable contribution from thermally activated parallel conduction in the GaN buffer was observed. To confirm this, Hall measurements were performed on a similar structure without the "back barrier", which showed no detectable changes in carrier density, indicating negligible buffer conduction.

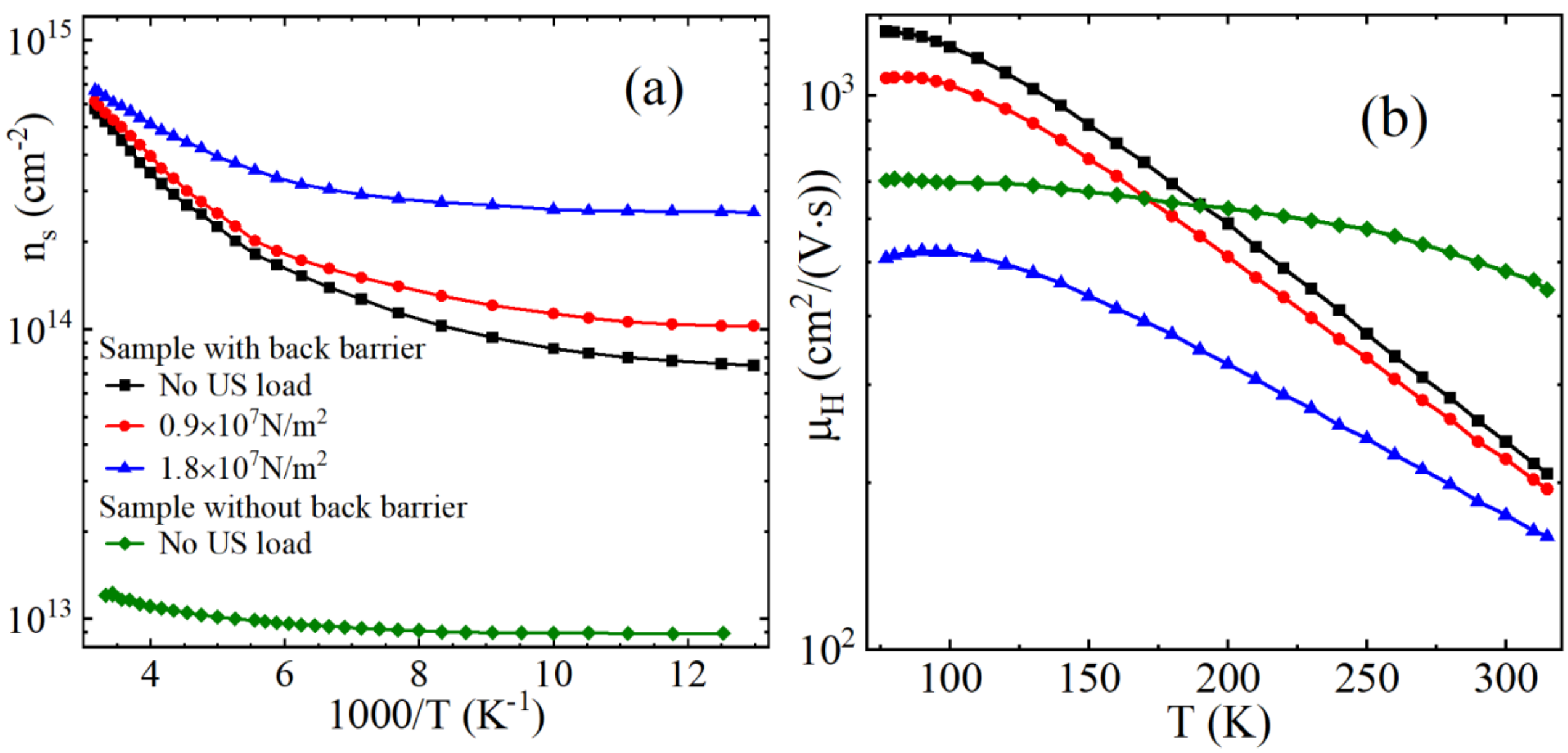

Fig. 2. Temperature dependence of the charge carrier concentration $n_s$ (a) and mobility $\mu_H$ (b) without and with applied ultrasonic loading ($f_{US}$ = 9 MHz).

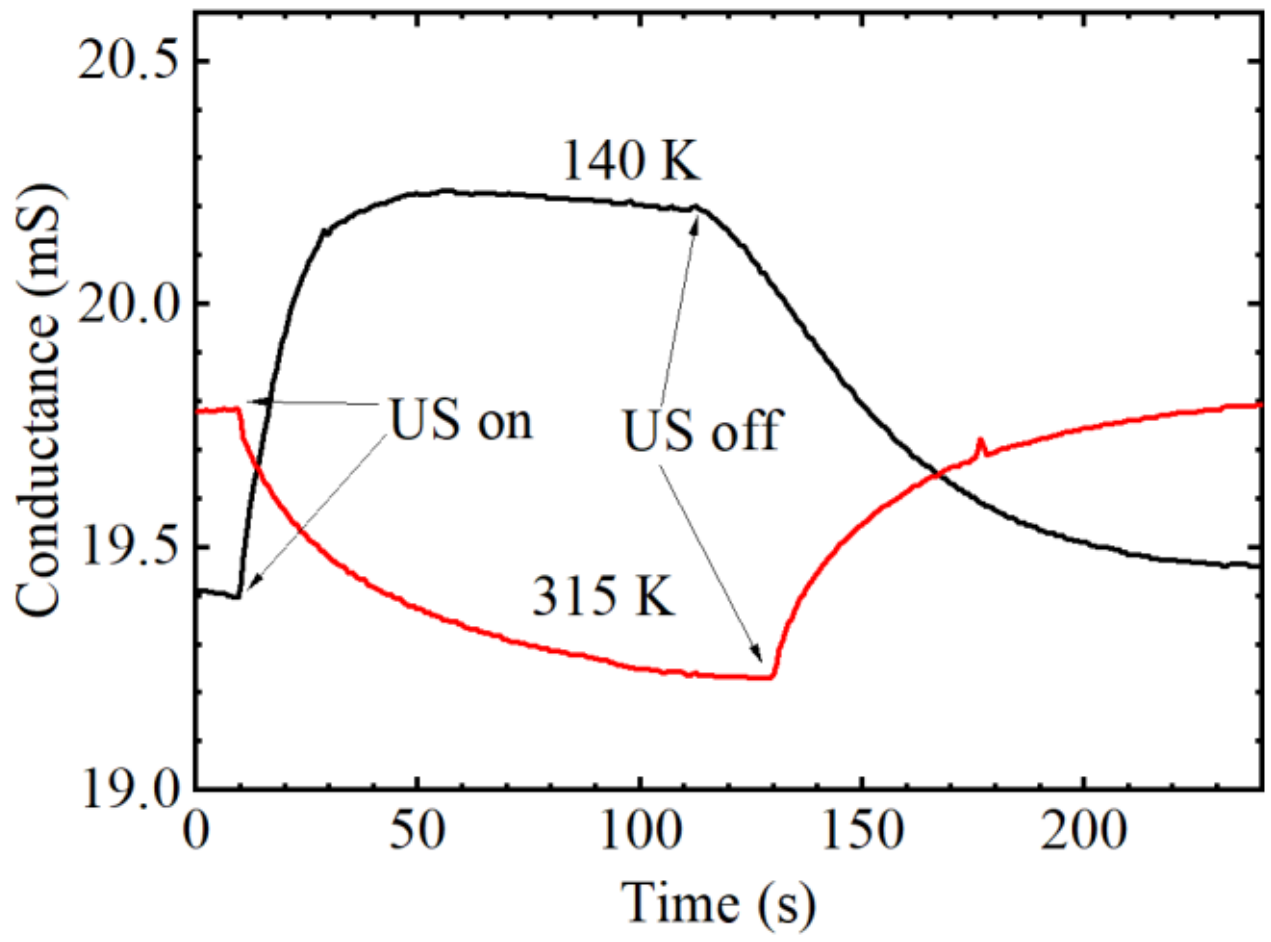

Fig. 3. Transient dependence of conductivity at different temperatures when ultrasonic loading of $\tau_{US}$=1.35×10$^6$ N/m$^2$ is switch-on or switch-off.

We observed that the conductivity in the 2DEG channel at the AlGaN/GaN interface is highly sensitive to ultrasound. Moreover, as shown in Fig. 3, acoustically induced changes in conductivity persist for a long period after the ultrasound is removed. This phenomenon, referred to as persistent acoustoconductivity (PAC), is analogous to persistent photoconductivity (PPC)[12], where the carrier density in the 2DEG channel is primarily increased due to the transfer of electrons excited from traps—such as DX centers—following their reconstruction under ultrasonic loading.[13,14]

## 3.1 Concentration changes from ultrasound application

One mechanism by which ultrasound can affect the concentration of charge carriers in the 2DEG is through changes in the piezoelectric charge induced by the acoustic field. The piezoelectric field arises from mechanical stresses between layers with different free-state lattice constants. In the absence of external influences, this results in a specific energy barrier for electrons in the conduction band, which determines the carrier concentration. When acoustic waves are applied, the resulting change in the piezoelectric field leads to a reduction in this energy barrier, which can be described by the following equation:

$$n=n_0 \exp(-(U_0-\gamma_n\tau_{US})/kT), \qquad (1)$$

where $U_0$ – activation energy under normal conditions, $U_{US}=U_0-\gamma_n\tau_{US}$ – activation energy under ultrasonic loading, $\tau_{US}$ – ultrasonic loading, $\gamma_n$ - a quantity that has the dimension of volume and is called the activation volume, and characterizes the process of transformation of charge carriers activated by the influence of acoustic waves. In the literature, this quantity is usually of the order of the volume where the acoustically induced charges are located.[15] At low temperatures, the number of charge carriers increases, which corresponds to the specified formula.

Another mechanism that can alter the charge carrier concentration is the rearrangement of metastable centers, specifically DX centers, located at the GaN/AlGaN interface. In A3B5-type semiconductor structures, DX centers are deep donor defects with metastable properties, leading to phenomena such as persistent photoconductivity and bistability in heterostructures, where recombination is particularly slow due to the AlGaN barrier.[16–18] Acoustic loading can induce positional changes of these DX centers, modifying their energy levels and reducing the energy barrier for electron excitation. Analysis of the temperature dependence of the charge carrier concentration yields activation energies from the slope of the curve at T>200 K: $E_0$=49.6 meV (no ultrasound), $E_1$=40.9 meV ($0.45\times10^6$ W/m$^2$), and $E_2$=26.9 meV ($1.35\times10^6$ W/m$^2$), indicating that ultrasonic loading lowers the activation energy for carrier excitation.

## 3.2 Amplitude dependence of electron concentration and mobility

Analysis of the amplitude dependence of the surface concentration $n(\tau_{US})$ and the mobility of charge carriers $\mu_H(\tau_{US})$ on the power of the applied acoustic load is necessary to understand the mechanism of ultrasound action in the structure. Analyzing the effects of both temperature and amplitude also allows us to better determine the scattering mechanisms and the action of ultrasound.

Fig. 4(a,b) shows the dependence of concentration and mobility on the power of the applied ultrasound for temperatures 140, 160, 195, 245 K. The changes in concentration and mobility exhibit different behaviors: $n(\tau_{US})$ shows a nonlinear dependence on the applied acoustic load, whereas $\mu_H(\tau_{US})$ decreases approximately linearly with increasing acoustic load, $\mu_H(\tau_{US})$~$1/\tau_{US}$, which is consistent with acoustic deformation of the sample lattice. To quantify the effect of ultrasound on the concentration, an approximation of the amplitude change was performed using the previously

mentioned formula (1). The coefficient $n_0$ can be calculated by taking the data at $\tau_{US}$=0.

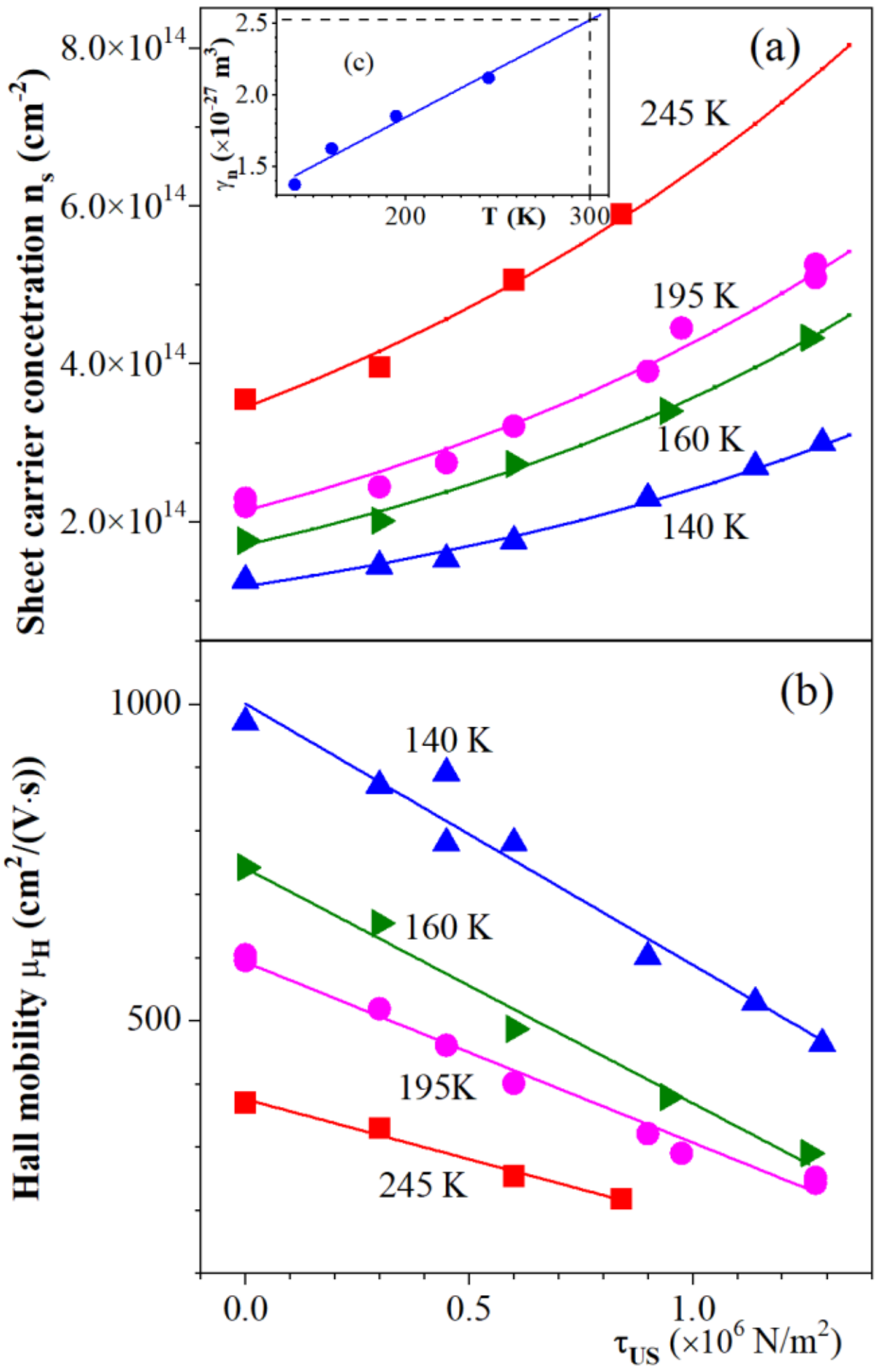

Fig.4. Dependence of the concentration (a) and mobility (b) of charge carriers on the applied field strength at different temperatures. Inset (c) demonstrates linear approximation of parameter $\gamma_n$ with approximation up to T = 300K according to the Table 1.

Fig. 4(c) shows an approximation of the value of $\gamma_n$. From the temperature dependence, we find that $\gamma_n(T=300\ K) \approx 2.53\times10^{-27}$ m$^3$. This value can be associated with the penetrating dislocations in the AlGaN layer, or, more precisely, with the near-dislocation volume. It has been shown early that acoustically induced rearrangement of point defects in semiconductors mainly occurs in the dislocation volume.[19] GaN/AlGaN heterostructures are characterized by a high concentration of dislocations that penetrate the AlGaN barrier layer. If we take the dislocation length to be of the order of the AlGaN layer thickness, $d$=20 nm, and the cross-section to be of the order of the square of the GaN lattice constant, $a$=3.19 Å, we obtain $V_{dis}\approx a^2\cdot d=2\times10^{-27}$ m$^3\approx\gamma_n$.

Using the obtained values of $\gamma_n$, we can calculate the change in potential $\Delta U_{US}=\gamma_n\tau_{US}$ for the amplitudes $\tau_{US}$=0; 0.45; 1.35×10$^6$ N/m$^2$. At a temperature of 300 K, $\gamma_n$=2.53×10$^{-27}$m$^3$ , corresponding to $\Delta U_{US}$=7.1 meV and 21.3 meV for the respective amplitudes. From the slopes of the experimental curves, the changes in activation energy are $\Delta E=E_0-E_1$=8.7 meV and $\Delta E=E_0-E_2$=22.7 meV. The close agreement between these values, with differences within 20% (considering experimental errors), supports the validity of the proposed model.

For mobility, which exhibits a linearly decreasing dependence on acoustic load, the following equation is used: $\mu_H(\tau_{US})=\mu_H(0)(1-c_\mu{}^\tau\tau_{US})$, where $\mu_H(0)$ is the mobility in the absence of load and $c_\mu{}^\tau$ is a proportionality coefficient that behaves similarly to $\gamma_n$. The approximation curves are presented in Fig. 4(b), and the corresponding parameters are listed in Table 1.

Table 1. Approximation parameters $n_0$ $(\tau_{US})$ and $\mu_H(\tau_{US})$

| T ,K | $n_o{}^{theor}$, $10^{14}$ $cm^{-2}$ | $n_o{}^{exp}$, $10^{14}$ $cm^{-2}$ | $\gamma_n$, $10^{-27}$ $m^3$ | $\mu_H(0)$, $m^2/(V\cdot s)$ | $c_\mu{}^\tau$, $10^{-3}$ $cm^2/N$ |
|---|---|---|---|---|---|
| 300 | | 5.2 | 2.525 | | |
| 245 | 3.88 | 3.445 | 2.117 | 376 | 2.53 |
| 195 | 2.55 | 2.143 | 1.848 | 592 | 2.41 |
| 160 | 1.8 | 1.709 | 1.622 | 741 | 2.51 |
| 140 | 1.4 | 1.185 | 1.372 | 1001 | 2.07 |

### 3.3 Diffraction scans of the structure under the influence of ultrasonic loading

The electrical properties of the heterostructure are primarily governed by deformation mechanisms arising from mechanical stresses between layers with mismatched lattice constants. To study the effect of acoustic loading, it is important to determine the amplitude of lattice oscillations (displacements), particularly in thin film structures.[20] In this work, X-ray diffraction experiments were conducted to investigate diffraction reflection curves (DRCs) under acoustic loading. The shape and amplitude of these reflection curves provide information about the structural quality and enable assessment of the influence of ultrasonic loading.

Fig. 5 shows the dependence of the relative change in the lattice parameter $\Delta C/C$ on the acoustic load for different frequencies of the piezoelectric transducer in the resonance region. On a logarithmic scale, a linear dependence of the logarithm $\ln(\Delta C/C)\sim\tau_{US}$ is observed. At a power of $\sim 2\times10^6$ $N/m^2$ and above, saturation of the effect is observed, especially at higher frequencies.

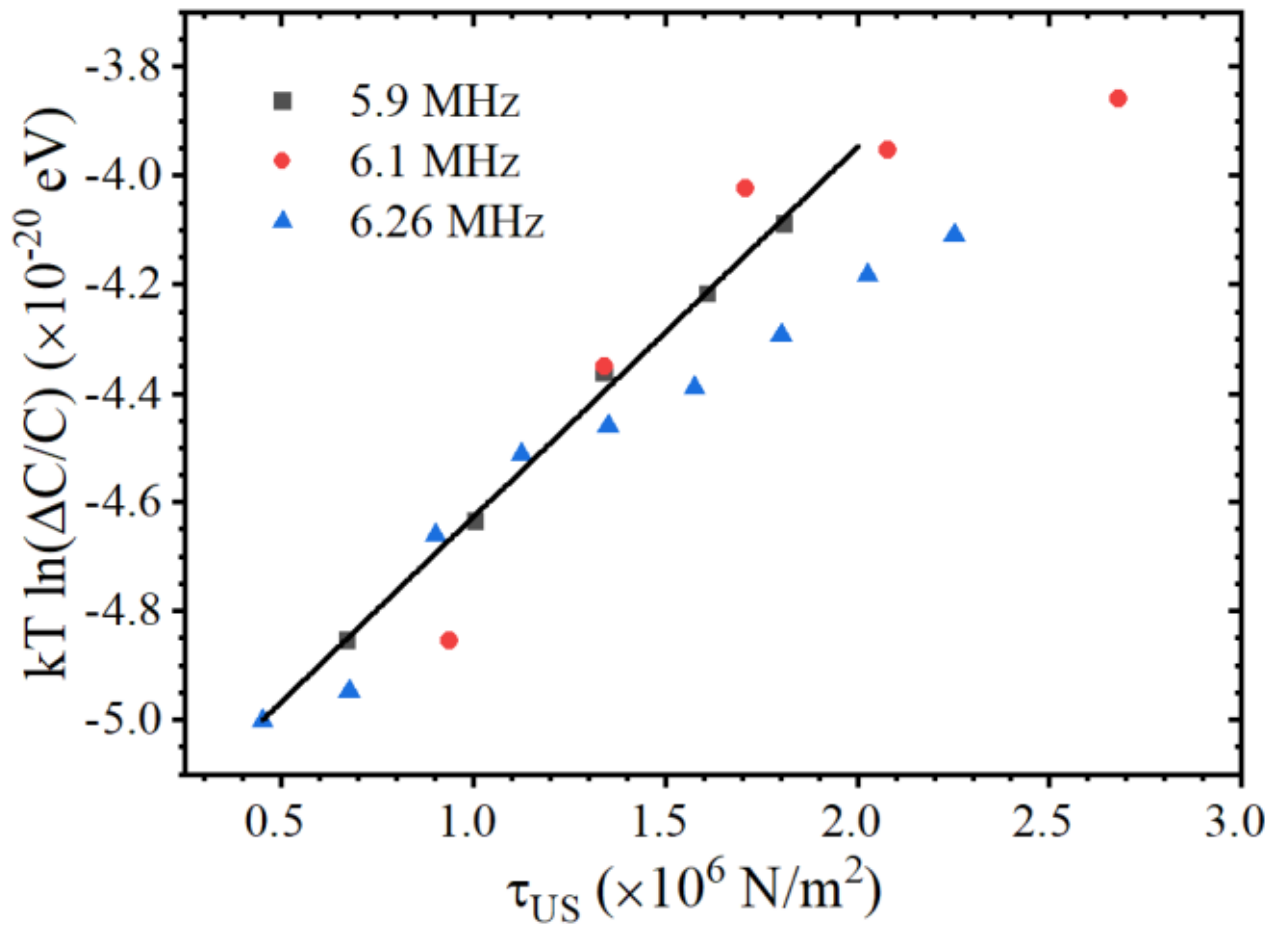

Fig. 5. Dependence of the logarithm of the lattice parameter $kT\cdot\ln(\Delta C/C)$ on the applied load $\tau_{US}$.

To explain the observed lattice expansion induced by acoustic loading, we employed a model based on acoustically active DX centers. In this context, we introduce the transition energy of the center, $U_{DX}$, and the activation volume, $\gamma_{DX}$, where the transition occurs. The influence of ultrasound on the lattice is described using a model analogous to the amplitude dependence of the charge carrier concentration $n(\tau_{US})$:

$$\Delta C/C = b\exp(-(U_{DX} - \gamma_{DX}\tau_{US})/kT), \tag{2}$$

It should be noted that this dependence is valid primarily in the low-load regime, as at higher loads, lattice expansion compensation effects become significant, leading to saturation of the effect. Additionally, the behavior varies with frequency. For example, at a frequency of 5.9 MHz, a linear approximation of the logarithmic dependence was performed:

$$kT\ln(\Delta C/C) = kT\ln b - U_{DX} + \gamma_{DX}\tau_{US} = U_{DX}* + \gamma_{DX}\tau_{US}, \tag{3}$$

where $U_{DX}*$ is the activation energy under ultrasonic loading, and is determined from a linear approximation, and $U_{DX}$ determines the transition energy in the absence of ultrasonic loading. The value of $b$ is considered independent of temperature, and its meaning can be explained as follows. Given a limited number of DX centers, at a certain power, the effect is depleted due to the completion of the reorganization of a significant number of centers from the $DX^-$ state to the neutral $DX^0$ states. Further expansion of the lattice occurs due to mechanical expansion from acoustic loading. This is especially noticeable for the frequency $f_{US}$=6.1 MHz at a power above $2.5 \cdot 10^6$ N/m$^2$. For the approximation, the maximum value of the parameter $b$ is determined to be $b \approx \max(\Delta C/C) \approx 10^{-4}$, which corresponds to $kT\ln b \approx 237$ meV.

### 3.4 DX Center Restructuring Model

Therefore, the values of the parameters $E_a$, $U_{DX}$, $\gamma_n$, and $\gamma_{DX}$, obtained from the experimental amplitude dependences of $n(\tau_{US})$ and $(\Delta C/C)(\tau_{US})$, are consistent with the proposed physical model.

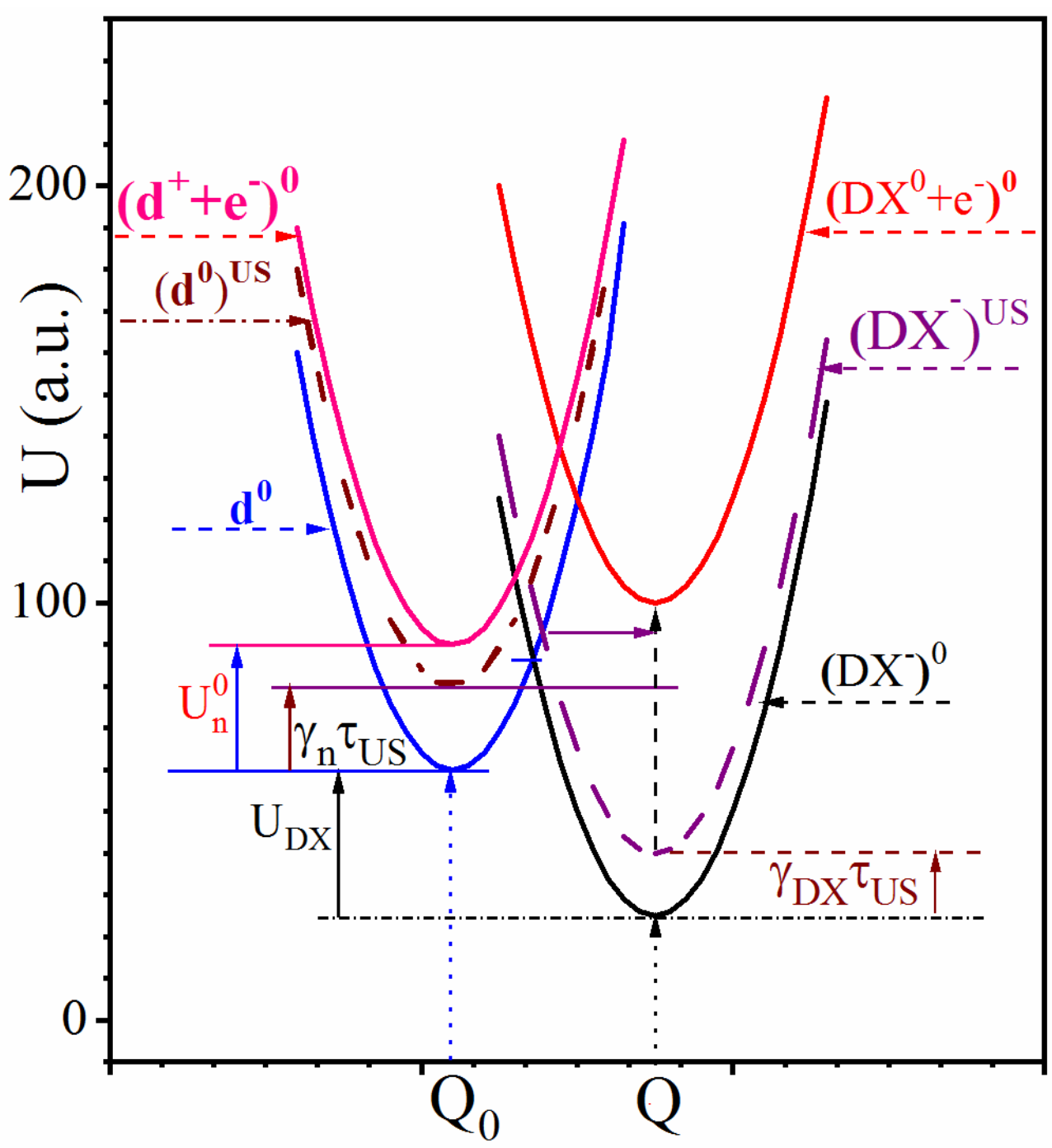

Fig.6. Energy model of acoustically induced rearrangement of a donor doubly charged metastable DX center.

We base our interpretation on the energy model of DX centers typical for heterostructures, as described in detail in Ref. [21], where various donor impurities can exist in the DX state. By adapting this model to account for ultrasonic loading, we can explain many of the observed acoustically induced effects. In the presence of a deformation ultrasonic field, the periodic modulation of the distance between possible donor atom positions reduces the barrier for transition to the DX state, resulting in an increase in both electron concentration and lattice parameter, as observed experimentally. Fig. 6 presents a generalized model of acoustically induced rearrangement of a donor doubly charged metastable DX center for $GaN/Al_{0.2}Ga_{0.8}N/AlN$ heterostructures. The corresponding energy characteristics under ultrasonic loading are denoted as ${U_n}^{US}$ and ${U_{DX}}^{US}$. The dashed lines represent the parabolic energy profiles for the doubly charged $(DX^-)^{US}$ state and the shallow donor $(d^0)^{US}$ state under ultrasonic loading. The acoustically induced shifts in ${U_n}^{US}$ and ${U_{DX}}^{US}$ are described by $\gamma_n\tau^{US}$ and $\gamma_{DX}\tau^{US}$, respectively. The origin of the traps is still under investigation, with possible sources including oxygen donors or compositional fluctuations in the $Al_xGa_{1-x}N$ alloy. Further work is ongoing to clarify the nature of these defects.

## 4. Conclusions

The effect of dynamic deformation on the transport characteristics of GaN/AlGaN heterostructures was investigated and the following conclusions were made:

1. After applying and removing an acoustic load, the heterostructures retain a stable conductivity similar to a stable photoconductivity. The sign of the change in conductivity also depends on temperature, and the magnitude of the change depends on the intensity of the applied load.

2. Applying a load leads to an increase in the concentration of charge carriers and a decrease in mobility, which also changes the conductivity of the sample.

3. From the amplitude characteristics of the sample, the characteristic energy of the center was obtained, which corresponds to the excitation under acoustic loading. A comparison was made with the energy of the center obtained from the changes in the constant lattice under the application of ultrasound.

4. To explain the change when applying dynamic loading, a model of a deep donor center (DX-center) with multi-stage restructuring was proposed.

## Acknowledgments

This study was supported by the National Academy of Sciences of Ukraine (Project No. 5.2/26-P). A.E.B. gratefully acknowledges the long-term program supporting the Ukrainian research teams at the Polish Academy of Sciences, which was carried out in collaboration with the U.S. National Academy of Sciences with the financial support of external partners (Project: LTP NAS KOCHELAP A7.11.0008; contract # PAN.BFB.S.BWZ.367.022.2023). The authors would also like to thank V.A. Kochelap and V.V. Koroteyev for the valuable collaborative work.

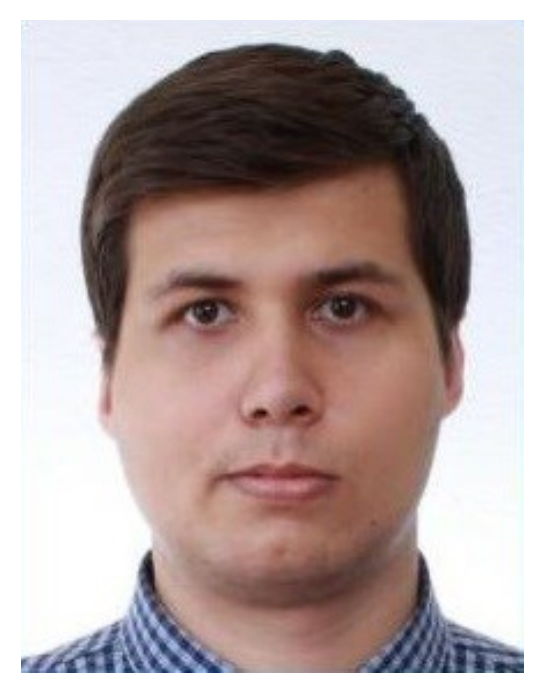

**Vladyslav Kaliuzhnyi** received his PhD degree at V. E. Lashkaryov Institute of Semiconductor Physics NAS of Ukraine in 2024, and works here as Junior Researcher. The area of scientific interests includes physics of semiconductor materials and devices (HEMTs and HEMT-like structures, III-nitrides), modeling of properties

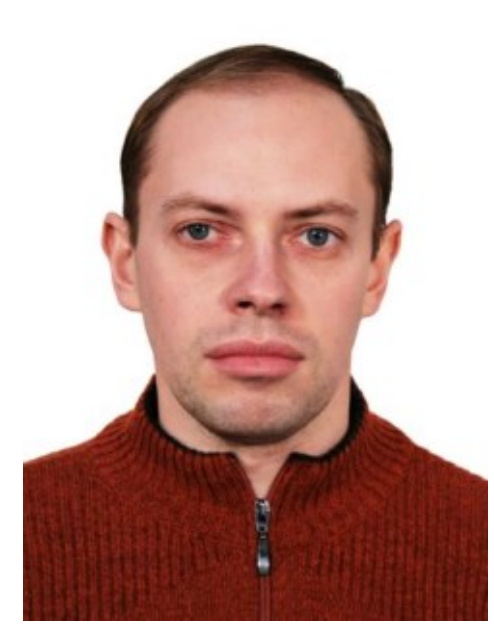

**Mykola Tymochko** received his PhD degree at V. E. Lashkaryov Institute of Semiconductor Physics NAS of Ukraine, and works here as Senior Researcher. His area of interest are transport and optical properties of semiconductor materials under influence of dynamic deformation.

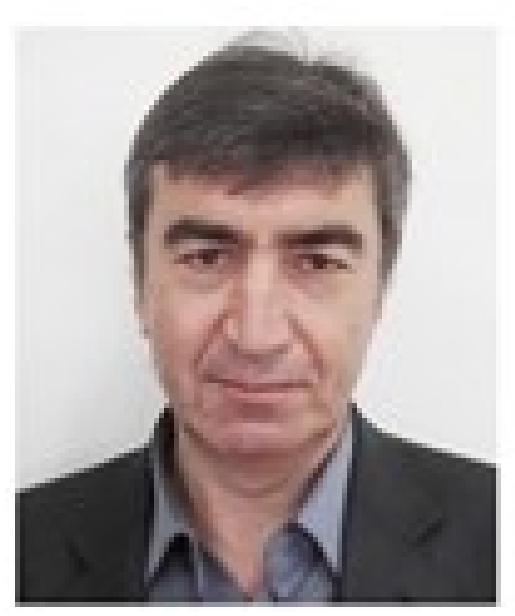

**Oleksandr Gudymenko**, got his PhD at V. E. Lashkaryov Institute of Semiconductor Physics, National Academy of Sciences of Ukraine (ISP NASU) in 2012. Now he is a Senior Researcher at the ISP NASU. Field of research: solid-state physics, dynamical theory of diffraction of radiation, X-ray optics, and X-ray diffraction analysis of semiconductor crystals.

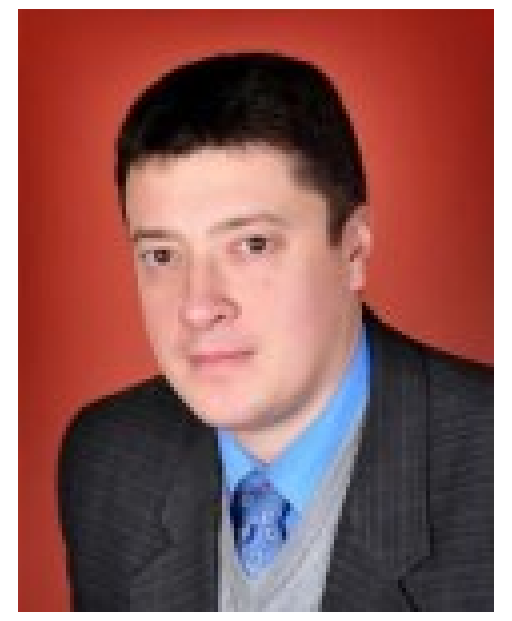

**Oleg Olikh** got his PhD degree in 2001 and Doctor of Sciences degree in 2018 at Taras Shevchenko National University of Kyiv (KNU). He is Professor and the Head of department of General Physics of KNU. His research areas include the effect of ultrasound on the substance, acoustic-stimulated dynamic phenomena in semiconductor barrier structures.

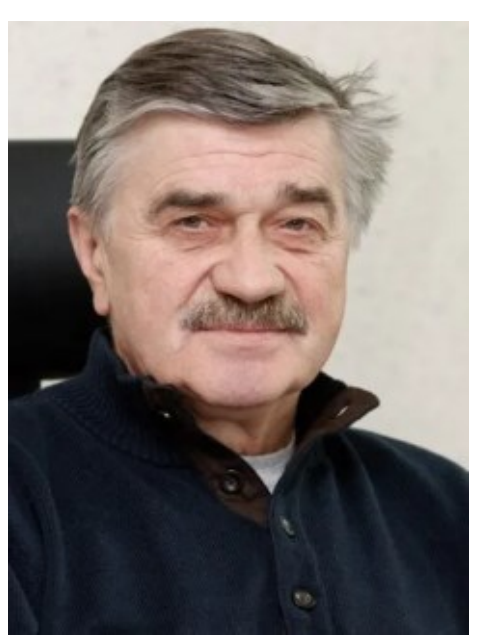

**Alexander Belyaev** got Doctor of Sciences degree at V. E. Lashkaryov Institute of Semiconductor Physics, National Academy of Sciences of Ukraine (ISP NASU) in 1991. Currently, he is NASU Academian, and Professor of ISP NASU. The area of his scientific activity is transport in quantum multilayer heterostructures and low-dimensional systems and their optical properties as well as application of such structures in UHF devices.